\def\arcsec{''}
\def\arcdeg{$^{\circ}$}
\def\arcmin{'}

\documentclass{ws-jai}
\usepackage[flushleft]{threeparttable}
\begin{document}

\catchline{}{}{}{}{} 

\markboth{P.C. Nagler}{Observing exoplanets in the near infrared from a high altitude balloon platform}

\title{Observing exoplanets in the near-infrared from a high altitude balloon platform}

\author{Peter C. Nagler$^{1}$, 
Billy Edwards$^{2}$, 
Brian Kilpatrick$^{3}$, 
Nikole K. Lewis$^{4}$, 
Pierre Maxted$^{5}$, 
C. Barth Netterfield$^{6}$, 
Vivien Parmentier$^{7}$, 
Enzo Pascale$^{8,9}$, 
Subhajit Sarkar$^{9}$, 
Gregory S. Tucker$^{3}$, 
Ingo Waldmann$^{2}$} 

\address{
$^{1}$NASA/Goddard Space Flight Center, Greenbelt, MD, USA\\
$^{2}$Department of Physics and Astronomy, University College London, Gower Street, WC1E 6BT, UK\\
$^{3}$Department of Physics, Brown University, 182 Hope Street, Providence, RI 02912, USA\\
$^{4}$Department of Astronomy and Carl Sagan Institute, Cornell University, 122 Sciences Drive, Ithaca, NY 14853, USA\\
$^{5}$Keele University, Staffordshire, ST5 5BG, UK\\
$^{6}$Department of Physics, University of Toronto, 60 St. George Street, Toronto, ON M5S 1A7, Canada\\
$^{7}$University of Oxford, Oxford OX1 2JD, UK\\
$^{8}$La Sapienza University of Rome, Rome, IT\\
$^{9}$School of Physics and Astronomy, Cardiff University, The Parade, Cardiff, CF24 3AA, UK
}

\maketitle

\corres{$^{1}$Peter C. Nagler, peter.c.nagler@nasa.gov}


\begin{center}
{\small Preprint of an article accepted for publication by the Journal of Astronomical Instrumentation, 2019.}
\end{center}

\begin{abstract}
Although there exists a large sample of known exoplanets, little data exists that can be used to study their global atmospheric properties. This deficiency can be addressed by performing phase-resolved spectroscopy -- continuous spectroscopic observations of a planet's entire orbit about its host star -- of transiting exoplanets. Planets with characteristics suitable for atmospheric characterization have orbits of several days, thus phase curve observations are highly resource intensive, especially for shared use facilities. In this work, we show that an infrared spectrograph operating from a high altitude balloon platform can perform phase-resolved spectroscopy of hot Jupiter-type exoplanets with performance comparable to a space-based telescope. Using the 
EXoplanet Climate Infrared TElescope (EXCITE) experiment as an example, we quantify the impact of the most important systematic effects that we expect to encounter from a balloon platform. We show an instrument like EXCITE will have the stability and sensitivity to significantly advance our understanding of exoplanet atmospheres. Such an instrument will both complement and serve as a critical bridge between current and future space-based near infrared spectroscopic instruments.
\end{abstract}


\section{Introduction}
\noindent At balloon altitudes ($\sim 40$ km), Earth's atmosphere is stable and nearly transparent; a balloon-borne telescope can observe from a space-like environment at a small fraction of the cost. Recent demonstrations of sub-arcsecond pointing stability and accuracy from a balloon platform \cite{stuchlik2015,romualdez}
enable balloon-borne telescopes to perform measurements that require both short and long term stability. 
One application of such a platform is to perform phase-resolved spectroscopic observations of hot Jupiter exoplanets in the near-infrared (NIR).

Phase-resolved spectroscopic observations are continuous spectroscopic observations of a transiting exoplanet as it orbits its host star. Such observations are rich scientifically. In addition to observing the two well-known discontinuities in a planet's orbit -- the eclipse and the transit -- observing the entire orbit enables constraints on the global energy budget and atmospheric circulation of exoplanets. By probing multiple wavelengths, and therefore pressures, these observations map the exoplanet's longitudinal heat distributions and vertical atmospheric structures. Considerable effort has gone into modeling the global circulation of hot Jupiters \cite[e.g.,][]{showman2002, cooper2005, langton2007, cho2008, showman2009, menou2009, dobbs-dixon2010, rauscher2010, lewis2010, heng2011, perna2012, showman2013, kataria2013, lewis2014}, but there is still very little spectroscopic observational data that can be used to study their global atmospheric properties. 

A balloon-borne instrument optimized for phase-resolved spectroscopy has the ability to substantially advance our understanding of hot Jupiter physics. A near-infrared spectrograph operating from a high altitude balloon platform can perform observations that are inaccessible from existing ground, airplane, or space-based observatories. Beyond atmospheric transparency and stability, a balloon platform allows continuous spectral coverage from 1\,--\,4 $\mu$m without contamination from atmospheric emission or emission from warm optics. The 1\,--\,4 $\mu$m range is rich with spectroscopic features of carbon and oxygen-bearing species like H$_{2}$O, CO, and CH$_{4}$, and probes a broad range of pressures in the atmosphere (1 bar to 1 mbar). Currently no space-based observatory can perform spectroscopic measurements across this entire range. The {\sl NIRSpec} prism on {\sl James Webb Space Telescope (JWST)} will cover the entire 1\,--\,4 $\mu$m range, but it will be limited to targets dimmer than J magnitude 9.83 \cite{Nielsen2016}. When deployed from the Arctic or Antarctic, many visible astrophysical targets never set, so long duration (several days or more), continuous observations are possible. There are only two spectroscopic phase curve measurements in the literature \cite{stevenson2014,Kreidbergetal2018}; the first required $\sim 60$ {\sl Hubble Space Telescope (HST)} orbits to measure a single exoplanet orbit, and the second required 46 hours of {\sl HST} time and 60 hours of {\sl Spitzer} time to do the same. The observed planets have orbital periods of under 22 hours. Phase curve measurements will be similarly resource-intensive for future space telescopes.

A balloon platform does introduce systematic effects that must be accounted for when designing an instrument optimized for phase-resolved spectroscopy. These include both short and long period altitude variations, variations in atmospheric transmission and emission as the telescope boresight changes, pointing jitter, field stop losses (if applicable), thermal variations in the optics, and detector variations. In this work, we quantify how these systematic effects impact observations of hot Jupiter exoplanets in the near infrared. Using the proposed EXoplanet Climate Infrared TElescope (EXCITE) experiment \cite{Tuckeretal2018} as a worked example, we demonstrate that residual systematic uncertainties due to the sources listed above are subdominant to photon noise for all reasonable integration times.

In Section \ref{sec:EXCITE}, we introduce the EXCITE experiment, review its science goals, and describe the instrument design. In Section \ref{sec:syserr} we discuss systematic effects that an instrument like EXCITE will experience. The discussion of systematic effects is divided into three parts. First, we treat uncorrelated errors (Section \ref{sec:syserr_uncorr}). Next we discuss correlated noise sources inherent to the balloon platform of an instrument like EXCITE (Section \ref{sec:syserr_corr}). Finally, we detail the implementation and results of an end-to-end simulation of EXCITE that demonstrates that an instrument like EXCITE is photon noise limited for all reasonable integration times (Section \ref{sec:sim}). Our conclusions are in Section \ref{sec:conc}.

\section{The EXCITE experiment}\label{sec:EXCITE}

\begin{figure}
\centering
\includegraphics[width=0.44\textwidth]{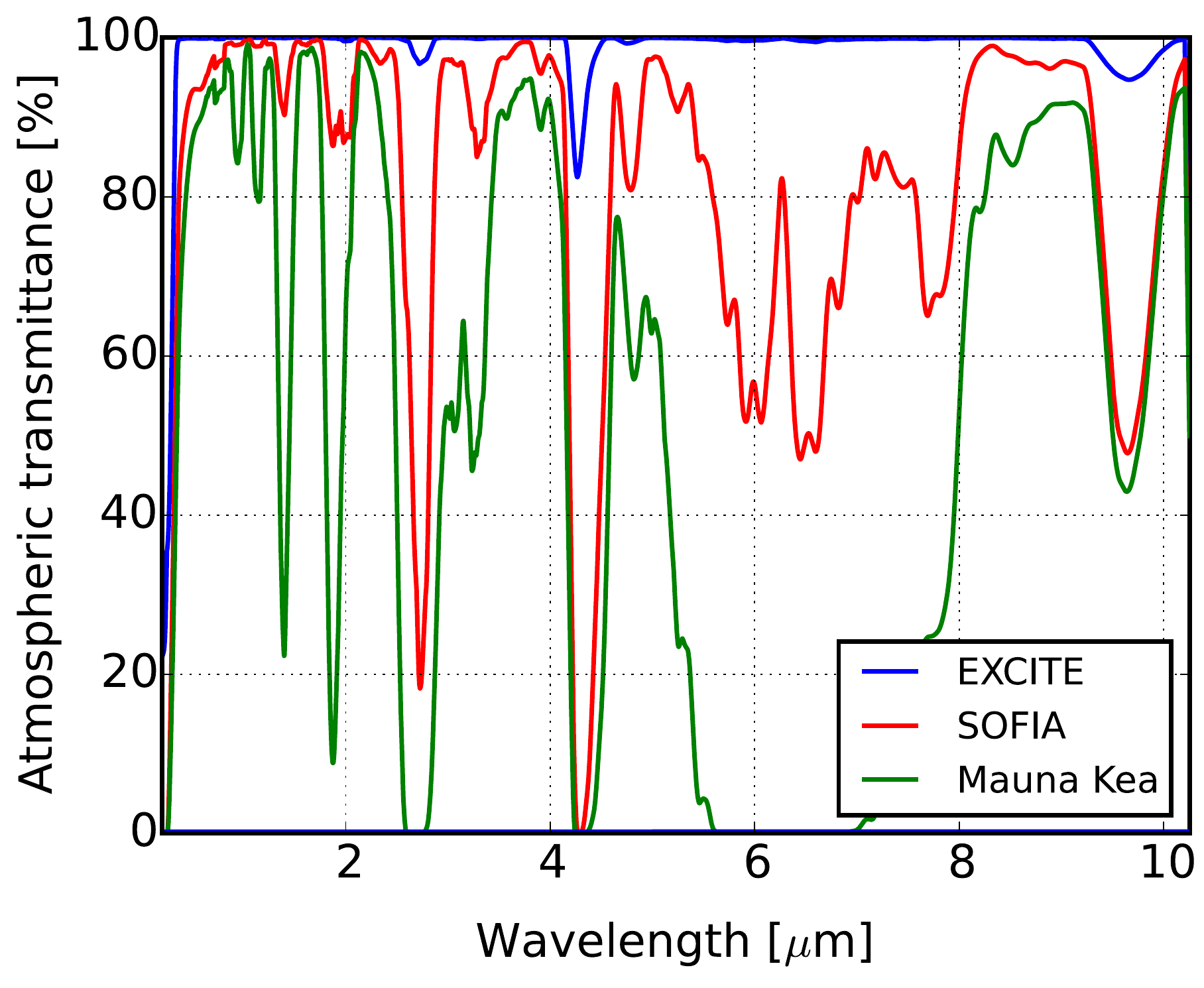}
\includegraphics[width=0.55\textwidth]{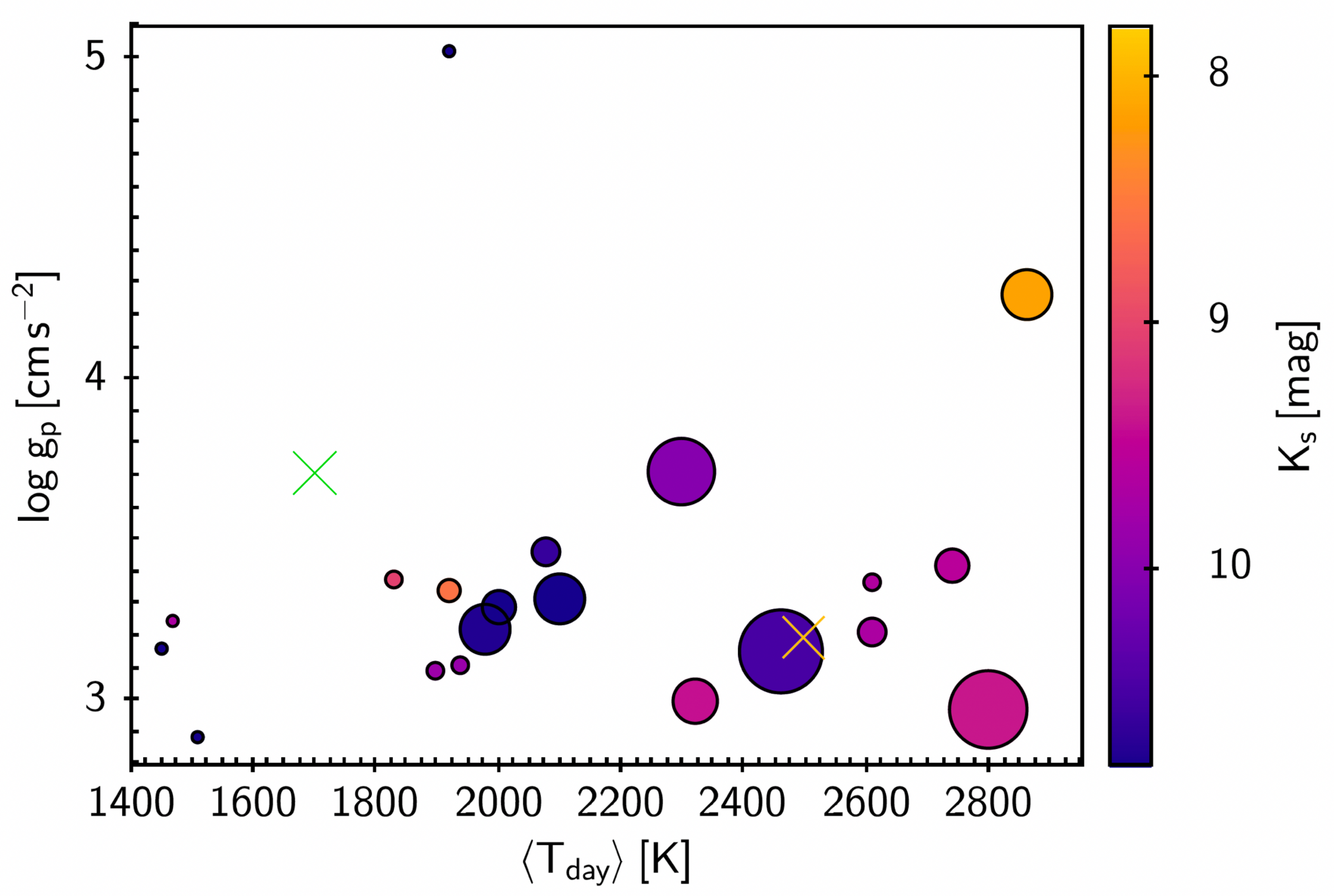}
\caption{Left: Atmospheric transmittance as a function of wavelength for observations from a LDB platform (blue), airplane (red), or the ground (green). EXCITE will observe in the band from 1--4 $\mu$m. Right: Properties of suitable targets for an Antarctic LDB flight. The mean dayside effective temperature $\langle {\rm T}_{\rm day}\rangle$ is estimated using an empirical relation calibrated on observed transiting hot Jupiters. These planets range in temperature from 1450 K to 2800 K. The symbol size is proportional to predicted K$_{\rm s}$-band eclipse depth, ranging from 260\,ppm to 2850\,ppm. Planets in this plot with low surface gravities ($\log {\rm g}_{\rm p}$) are also suitable targets for study with transmission spectroscopy during transits. The crosses show the properties of WASP-43b and WASP-103B, the only planets for which phase-resolved spectroscopy has been published to date \cite{stevenson2014,Kreidbergetal2018}.}
\label{fig:EXCITE_transmittance}
\end{figure}

The EXCITE experiment \cite{Tuckeretal2018} is a proposed balloon-borne spectrograph designed to perform phase-resolved spectroscopy of transiting hot Jupiter-type exoplanets. With moderate resolving power ($R \sim 50$ at the Rayleigh criterion) and continuous spectroscopic coverage across wavelengths from 1--4 $\mu$m, EXCITE will make observations that are inaccessible from existing observatories. It will help to bridge the gap between current (e.g., {\sl HST}) and future (e.g., {\sl JWST}) space-based NIR spectroscopic observatories. EXCITE will operate from a high-altitude ($\sim 40$ km) long duration balloon (LDB) platform, observing from a near-space environment above $>99\%$ of the Earth's atmosphere. At these altitudes the Earth's atmosphere is stable and nearly transparent. This reduces the impact of systematic effects due to atmospheric variations that can limit lower-altitude experiments (see Figure \ref{fig:EXCITE_transmittance}). EXCITE's $0.5$ m optical telescope assembly (OTA) is based on the successful Balloon-borne Imaging Telescope (BIT) platform~\cite{romualdez}. The pointing accuracy and stability of the BIT platform, combined with the circumpolar orbit of an Antarctic LDB flight, allows EXCITE to continuously stare at targets through the duration of their orbits (up to several days). Such observations are resource-intensive for shared-use facilities, making them well suited for a purpose-built platform. 

\subsection{Scientific motivation}

The orbit of a transiting exoplanet about its host star produces a time-varying signal in the flux measured by a distant observer. This signal is some small fraction of the planet-to-star flux ratio $\rho$ (generally $\rho<10^{-4}$). There are two well-studied discontinuities in the signal that yield significant insights into exoplanet atmospheric physics. The first is the transit, which occurs when the planet passes between an observer and its host star. During this event, starlight is filtered by the planet's atmosphere. Measurements of the resulting transmission spectrum can be used to constrain the planet's atmospheric composition. The information gained from transit measurements is limited, however. Transmission spectra only probe the low pressure atmospheric region near the limb of the planet. The second discontinuity is the eclipse, which occurs when the planet passes behind its host star as seen by an observer. Eclipse measurements can be used to obtain the planet's emission spectrum. The emission spectrum can provide information about atmospheric composition, vertical thermal structure, and circulation processes.  Still, eclipse observations provide only a hemispheric average of only the planet's dayside properties. 

Additional information can be gained by continuously observing the entire orbit (``phase curve'') of the planet, including the eclipse and transit. Phase curve measurements probe the emitted flux of a planetary atmosphere across all longitudes, enabling the study of how atmospheric properties vary with temperature for a fixed elemental composition and gravity. Phase curve measurements are rich scientifically. For example, the amplitude of phase curve variations measured at infrared wavelengths constrain the efficiency with which the planet recirculates energy from its permanent day side to its night side \citep[e.g.,][]{cowan2011b}.  The phase of the the phase curve's peak measures the offset of the brightest point on the planet from the sub-stellar point and indicates the relative strength of radiative and dynamical timescales in the planet's atmosphere \citep[e.g.,][]{showman2009, cowan2011a}. The overall shape of the phase curve can reveal chemical gradients in the planet's atmosphere \citep[e.g.,][]{knutson2012, lewis2014}.

\begin{figure}
\centering
\includegraphics[width=0.9\textwidth]{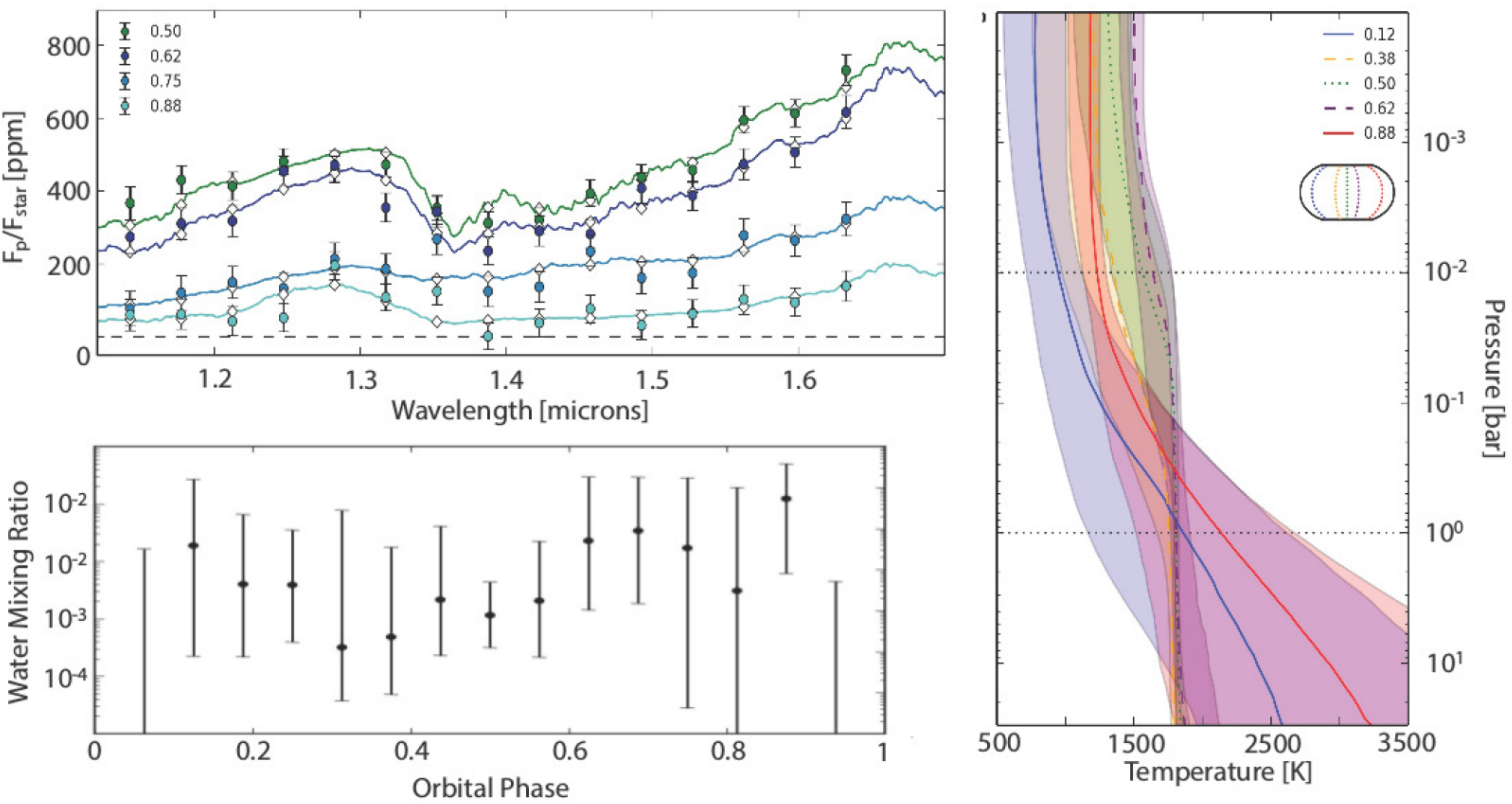}
\caption{Illustration of the power of spectroscopic phase curves. These results, for the hot Jupiter WASP-43b, are one of only two published spectroscopic phase curves to date~\cite{stevenson2014}.  The top left panel shows the {\sl HST}/WFC3 spectra for 4 different phases.  The solid lines represent model fits to the data at each phase (diamonds with error bars). The right panel shows the thermal profile constraints derived from a few select phases. These constraints are strongest between the horizontal dotted lines. The bottom left panel shows how the water abundance varies with phase.  While the error bars are large, this nearly uniform pattern with phase is consistent with predictions from coupled chemical-dynamical models~\cite[e.g.,][]{cooper2005, Agundezetal2014}.} 
\label{fig:Phasecurve}
\end{figure}

Phase curve observations are especially powerful when they are simultaneously performed in multiple spectroscopic bins (``phase-resolved spectroscopy'', Figure \ref{fig:Phasecurve}). Spectroscopic phase curve observations probe a wide range of pressure levels in the planet's atmosphere as the atmosphere's opacity changes as a function of wavelength.  Comparisons between phase curves measured at a range of wavelengths reveal how the relevant radiative, chemical, and dynamical timescales vary as a function of atmospheric pressure and therefore altitude. Spectroscopic phase curves measured through the peak of the planet's spectral energy distribution (typically at a wavelength of $\sim 2$ $\mu$m for hot Jupiters) reveal how the chemistry and dynamics of exoplanet atmospheres vary with equilibrium temperature and surface gravity. These are the only observations that directly constrain the global energy budget and circulation patterns that result from the extreme irradiation of hot Jupiters.

Recent work \cite[{\sl e.g.},][]{Arcangelietal2018,Kreidbergetal2018,Lothringeretal2018,Parmentieretal2018,Kitzmannetal2018,BellandCowan2018,KomacekandTan2018} has shown that many molecular species ({\sl e.g.,} H$_{2}$, H$_{2}$O, TiO, VO) should dissociate on the planet day-side for very hot planets (T$_{\mathrm{eq}}\sim 2200$ K), but recombine on the night-side. H$^{-}$ becomes an important opacity source on the day-side between 1 and 1.5 $\mu$m. As a consequence, the 1\,--\,2 $\mu$m day-side spectrum closely resembles a blackbody spectrum with a reduced water feature due to dissociation and presence of H$^{-}$ opacity in the water gaps. This limits the information that can be derived from short wavelength spectra. One molecule that does {\it not} dissociate until even higher temperatures is CO. As a result, at wavelengths longer than 2 $\mu$m there are strong features to be observed in emission spectra: the CO feature at $\sim 2.5$ $\mu$m and the resurgence of H$_{2}$O because of the higher opacity and the lack of absorption from H$^{-}$. This highlights the importance of {\it broadband} spectroscopic phase curve observations in the NIR.

For a typical LDB flight duration of $\sim 20$ days, and for an average orbital period of $\sim 2$ days, EXCITE can measure phase curves of up to ten planets in a single flight, greatly expanding our understanding of hot Jupiter atmospheric physics. Figure \ref{fig:EXCITE_transmittance} shows an example exoplanet physical parameter space that can be probed from an Antarctic LDB flight. Shown are relevant published parameters of known exoplanets that are both visible from the Antarctic and which never set. The magnitude of the host star largely determines the signal-to-noise ratio (SNR) of observations. The eclipse depth determines the sensitivity required in order to measure the phase curve amplitude.

\subsection{Instrument design}

In this section we describe the EXCITE instrument design. We pay particular attention to the optics, focal plane, and pointing system. The nominal properties of these components are described here. Non-idealities, and their impact on EXCITE's observations, are treated in Section \ref{sec:syserr}.

\subsubsection{Optics}
The EXCITE OTA is based on the telescope used successfully in the first test flight of BIT~\cite{romualdez}.  The only significant difference is that the BIT telescope included field correction lenses.  Since EXCITE is not an imaging instrument, distortion-free images across the focal plane are not required and image corrector lenses are not needed. The telescope has a 450 mm back focal length at $f/10$. The telescope structure is made from materials with similar coefficients of thermal expansion (CTEs) to ensure optical stability.  In addition the telescope can be focused during flight by moving the secondary mirror. This was successfully demonstrated in the BIT flight.

\begin{figure}
\centering
\includegraphics[width=0.49\textwidth]{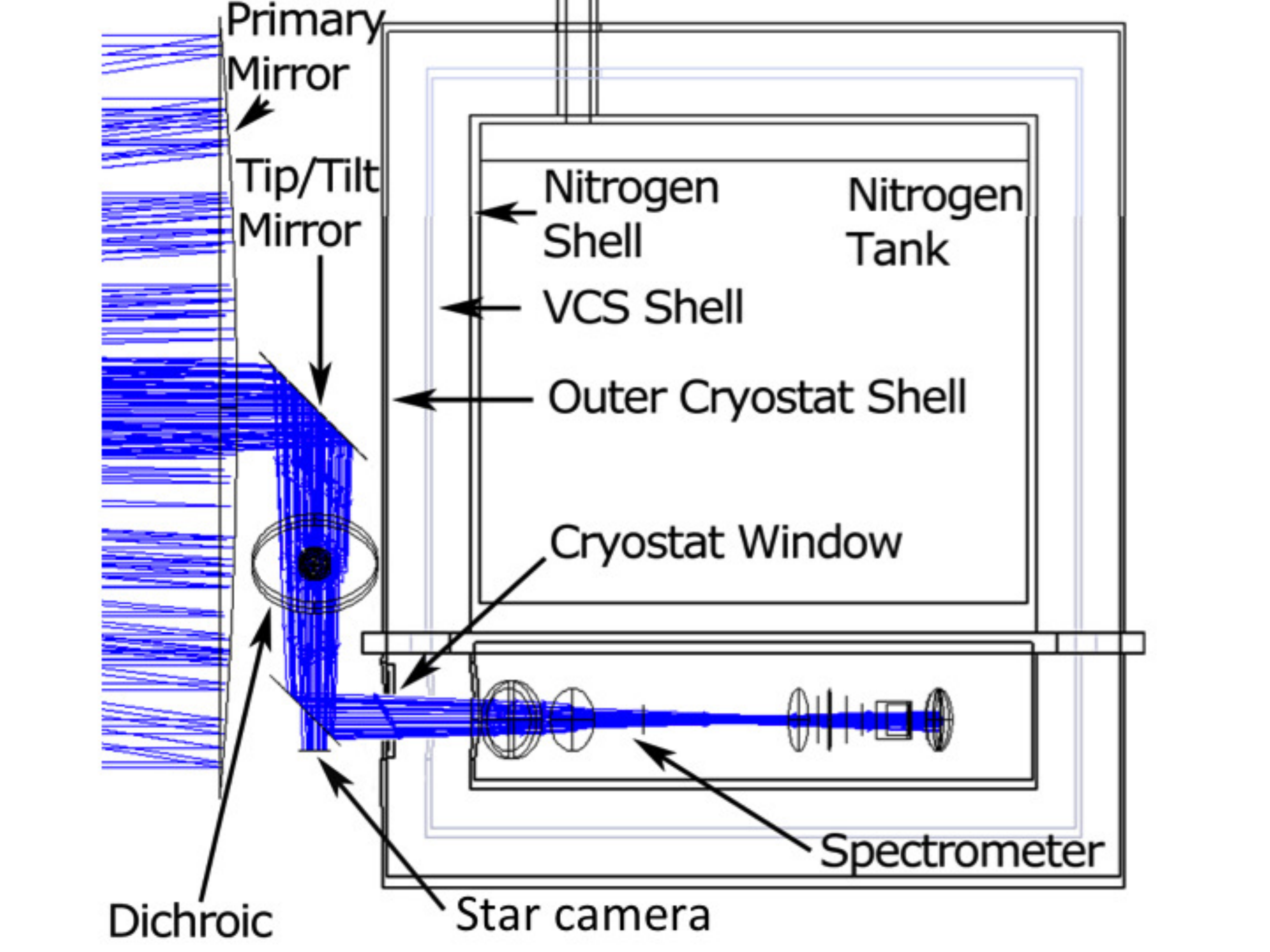}
\includegraphics[width=0.49\textwidth]{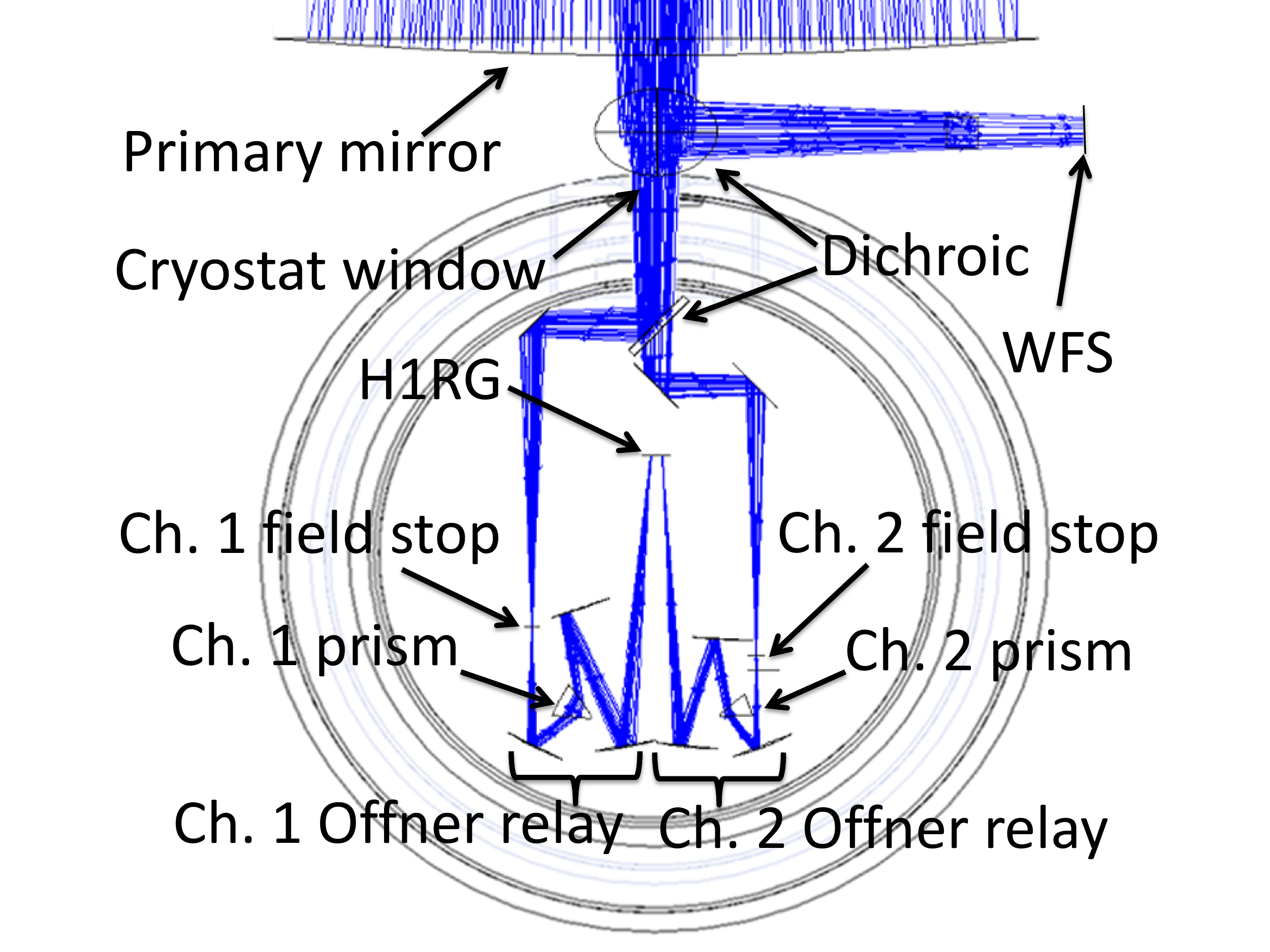}
\caption{Design of the optical system.  Left: Side view of the spectrometer showing the fine pointing tip/tilt mirror and the first dichroic. The real-time star camera is also visible. The secondary mirror is not visible. The spectrometer is located inside the solid nitrogen-cooled cryostat.  Right: Top view of the spectrometer. Also visible is the wavefront sensor.}
\label{fig:spectrometer_zemax}
\vspace{-0.5cm}
\end{figure}

Figure \ref{fig:spectrometer_zemax} shows the basic optical layout of the instrument and a Zemax ray trace of the optics. After passing through the telescope, the light first encounters an ambient temperature dichroic, which reflects wavelenths from 650\,--\,1000 nm and transmits wavelengths from 1\,--\,5 $\mu$m. The reflected light then encounters a second ambient temperature dichroic that reflects wavelengths from 650\,--\,850 nm and transmits wavelengths from 850\,--\,1000 nm. The 650\,--\,850 nm band is used by the real time star camera, which provides feedback to the tip-tilt mirror for the fine guidance system (Section \ref{sec:pointing}).  The 850\,--\,1000 nm band is used by the wavefront sensor (WFS).

The telescope secondary focus is used to focus light on the spectrometer field stops. The field stops are used to limit the field of view of the detector, reducing flux due to atmospheric and instrumental emission. The real-time star camera and WFS assembly are mounted on a separate focus mechanism so that they can be focused independently of the spectrometer. The star camera/WFS assembly will use a focus mechanism similar to that used for BIT.

The beam transmitted by the first dichroic passes into the cryostat through a CaF$_2$ window.  All the optics inside the cryostat are cooled to 62 K using solid nitrogen.  Inside the cryostat, the beam is split by a third dichroic into a reflected 1\,--\,2 $\mu$m band (``channel 1'') and a transmitted 2\,--\,5 $\mu$m band (``channel 2''). At the telescope focus, each beam passes through a field stop optimized for the respective bandpass. The width of each field stop is equal to the Airy diameter at a wavelength of 2 $\mu$m in channel 1 and a wavelength of 4 $\mu$m in channel 2 (50 $\mu$m and 100 $\mu$m wide, respectively). 

After the field stops, each beam passes through an Offner relay, with prisms positioned after the first concave mirrors. The mirrors in each Offner relay are gold-coated for a flat reflectivity from the visible to $\sim20$ $\mu$m. They yield an $f/10$ beam in channel 1 and an $f/15$ beam in channel 2. At the focal plane, the channel 1 beam is spread over 130 pixels, each of width 0.75\arcsec. The channel 2 beam is spread over 200 pixels, each of width 0.50\arcsec. The point spread function (PSF) of each beam is Nyquist sampled. The channel 1 beam is slightly defocused to achieve the desired sampling. Along the slit direction, the detector's full 1024 pixel range (Section \ref{sec:FPA}) will allow us to obtain sky background spectra (within the field stop) over an 8.6 arcmin field in channel 1 and a 12.7 arcmin field in channel 2.

The only non-reflective optical components of the OTA/spectrometer assembly are the ambient temperature (first) dichroic, the cryostat window, the cold (third) dichroic and the prisms.  Each of these components will have mean in-band transmittance exceeding 85\%. As a result, the overall optical efficiency will be greater than 60\%. With near-diffraction-limited performance over the bandpass, the mean spectral resolution (Rayleigh criteria) is $R= \lambda/\Delta\lambda \sim 50$. 

\subsubsection{Focal plane and calibrator}\label{sec:FPA}

The focal plane consists of a Teledyne HAWAII-1RG (H1RG) $1024\times 1024$ pixel detector array (18 $\mu$m pixel pitch), read out by a Leach controller \cite{LEACH}. The H1RG array has $\ge 85$\% quantum efficiency in the 0.6--5.3 $\mu$m band, with $\ge 95$\% pixel operability and correlated double sampling (CDS) read noise of $\le 15$ e$^{-}$ RMS. The instrument has sensitivity up to 5 $\mu$m, but beyond $\sim 4$ $\mu$m emission from the warm optics and atmosphere dominate the signal. The detector will be maintained at an operating temperature just above 62 K, where the median dark current level is $\le 0.5$ e$^{-}/{\rm s}$ per pixel~\cite{Chuhetal2006} in the science band (1\,--\,4 $\mu$m). 

An internal calibrator is used to monitor the performance of the detector array over time, correcting for detector system gain variations. The thermal source is the same as that used on {\sl JWST/MIRI}: a wound tungsten coil, spot welded with copper-clad nickel-iron core alloy, mounted and calibrated at Cardiff University. The filament can be driven with currents in the range from 0\,--\,15 mA and produces a maximum temperature of  $\sim1300$ K (with a power dissipation of 38 mW). Light from the calibrator is injected through a central hole in the first parabolic mirror of each Offner relay (see Figure \ref{fig:spectrometer_zemax}). This hole is co-aligned with, but smaller than, the telescope central obscuration and allows sufficient optical power to illuminate the array. This approach yields high flat field signal-to-noise when the calibrator source is driven to a temperature of $\sim1000$ K. This calibrator was extensively studied in the context of the ESA M3 candidate transit spectroscopy space mission, EChO, and M4 ARIEL. The study \citep{echoyb} demonstrated that this space-qualified source and its driving electronics can provide a reference signal to monitor flat field variations. This process is used to decorrelate photometric uncertainties from pointing jitter (Section \ref{sec:syserr_corr}).

The H1RG detector allows for windowing, where subsections of the array are read out by separate channels. This enables different exposure times for different sub-array regions. Without windowing, a maximum exposure time of $\sim 10$ ms would be required to avoid saturation of pixels sampling wavelengths longer than the $\sim 4.3$ $\mu$m atmospheric emission feature (see Figure \ref{fig:noise_uncorr}). Such short integration times would result in a large read noise contribution to the error budget at the red end of the 1--2 $\mu$m band, where the faintest stars allow for exposures of up to 50 s. These long exposures can also be achieved by up-the-ramp sampling, allowing for saturation of pixels sampling the 4\,--\,5 $\mu$m region of the spectrum when desired. Reading detector pixels up-the-ramp also allows for monitoring and rejection of cosmic ray events \cite{Offenbergetal2001}.

\subsubsection{Gondola and pointing}\label{sec:pointing}

The EXCITE gondola and pointing system are based on the BIT platform, which was successfully demonstrated in a 2015 flight~\cite{romualdez}.  From the 8 hour flight, sub-arcsecond pointing stabilization of the telescope frame was demonstrated over tracking periods of more than an hour at a time.  During a given tracking period, the telescope was stabilized to approximately 0.05\arcsec~RMS~\cite{romualdez}.

\begin{figure}[t]
\centering
\includegraphics[width=1.0\textwidth]{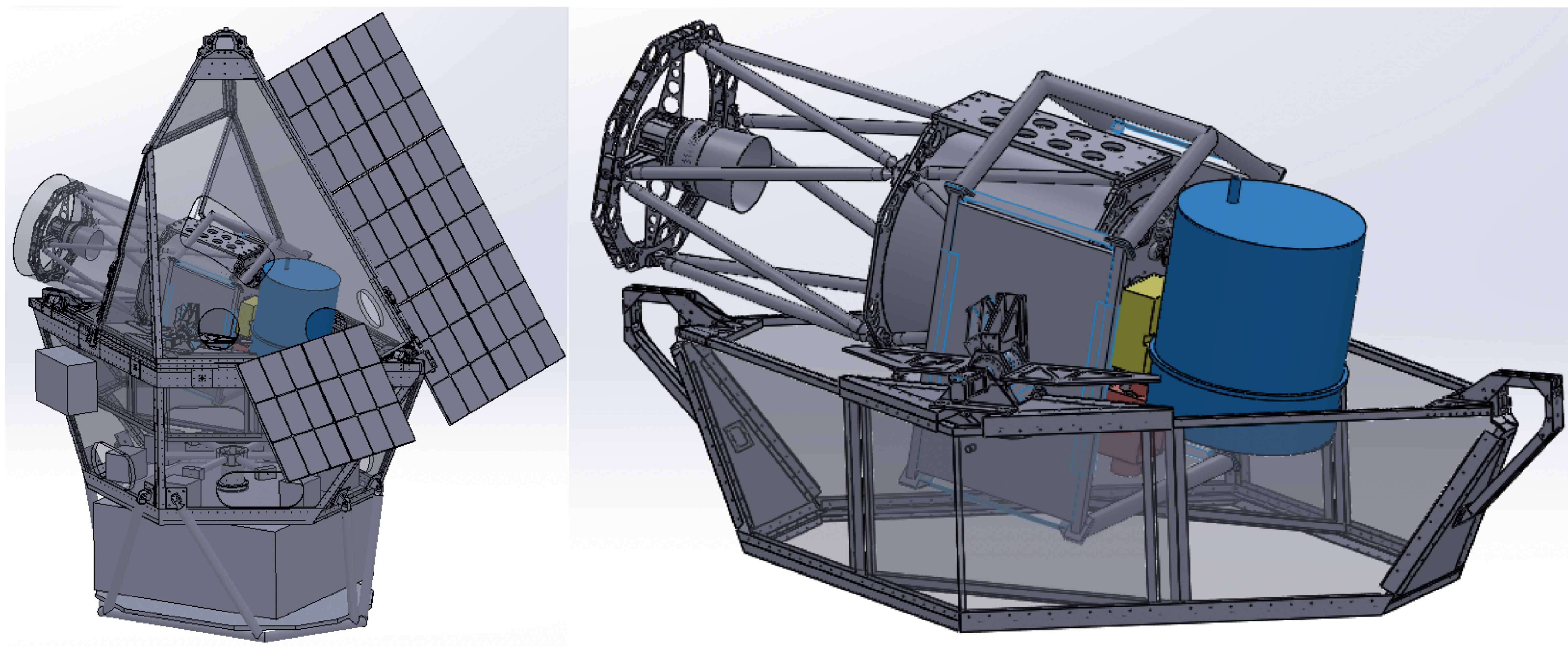}
\caption{The EXCITE instrument.  The telescope is a Ritchey-Chr\'etien design with a 0.5 m diameter primary mirror, which is pointed to an accuracy of 2\arcsec~in all three axes.  A fine guidance tip/tilt mirror provides the required 0.1\arcsec~pointing stability.  The telescope will be surrounded by a temperature stabilized shroud.  BIT successfully used such a shroud stabilized to $\pm 100$ mK.  The temperature is controlled by electrical heater pads.  For the entire LDB flight, EXCITE can perform with the same temperature stability demonstrated by BIT.\label{fig:gondola}}
\end{figure}

The EXCITE gondola (Figure \ref{fig:gondola}) is designed to hold the 0.5 m diameter telescope in three axes with a peak-to-peak accuracy of better than 2\arcsec.  Pointing to this precision is challenging.  Accelerations of the balloon produce motions of the outer frame of around 6\arcmin~in both pitch and roll. Additionally, the balloon rotates at a characteristic rate of $\pm 0.5$\arcdeg/s.  The motion of the inner frame needs to be decoupled from these, without introducing any high frequency shocks from static bearing friction and without driving gondola vibrational modes. EXCITE achieves this by actively controlling the orientation of the telescope in three nearly orthogonal axes with feedback from a series of pointing sensors.

Yaw (approximately azimuth) is controlled by a high-moment reaction wheel and an actuated pivot which torques against the flight train. Torque discontinuities due to static friction are avoided by biasing the average reaction wheel rotation rate to 15\arcdeg/s so that it never stops while pointing.  To avoid torque discontinuities with the pivot, which will routinely go through zero speed, the pivot is driven by a microstepped stepper motor geared down with a 100:1 gear reducer.  This allows smooth motion at any balloon rotation speed.

The roll and pitch axes are supported through flexure bearings and driven by brushless DC motors.  The torque from the flexure bearings, which are essentially rotational springs, is continuous, so there are no torque discontinuities in either the roll or pitch axis during a pointing observation.  The flexure bearing rotation is limited to $\pm 6$\arcdeg.  For an Antarctic LDB this range of motion allows EXCITE to track the sky for at least 4 hours and up to $\sim 8$ hours or more at a time. This time assumes typical Antarctic LDB latitudes between 73$^{\circ}$ and 82$^{\circ}$ S.  About 30 seconds are required to reset the flexure after it reaches its limit of motion, so the operating conditions of the telescope will not change noticeably.

As the nitrogen in the cryostat evaporates during flight, the balance of the inner frame will shift.  To compensate for this a balance system will be periodically adjusted to bring the inner frame back into balance.  The balance system works by driving a mass along the telescope boresight outside of the baffle.  Balance is monitored by observing motor current required for pointing.  

Coarse elevation control is achieved using a set of conventional ball bearings in series with the pitch axis flexure bearing, driven by a stepper motor.  The stepper motor and ball bearings are locked at the target elevation throughout a pointing integration.  The telescope pitch range is 22$^\circ$ to 57$^\circ$.

Coarse azimuthal pointing is provided by a magnetometer, which will give attitude information good to several degrees, and a set of pin-hole sun sensors, similar to those developed for BLASTpol \cite{Korotkov2013} and Spider, which will provide pointing to 0.5 degrees. These course pointing systems are sufficient to verify the lost-in-space solutions from the CCD cameras.  A pair of SBIG CCD cameras with 680 nm low pass filters, similar to those used on the Spider payload, will provide multi-star images, which will allow for lost-in-space solutions at a 0.5 Hz rate. These cameras are placed orthogonal to each other to provide a full characterization of the inner frame orientation. The camera sensitivity is adequate to provide several stars in any random pointing.

Fine pointing control, at the 0.05\arcsec~level, is provided by a piezoelectric-driven tip/tilt-Platform before the focal plane.

\section{Systematic effects}\label{sec:syserr}

In this section, we examine the most important systematic effects we expect to encounter while observing exoplanets from a LDB platform in the NIR. We treat systematic effects in several ways. In Section \ref{sec:syserr_uncorr}, we develop a radiometric noise model that accounts for the known uncorrelated noise sources. These include photon shot noise from the target exoplanet's host star, the Earth's atmosphere, and the telescope optics, and detector dark current and read noise. From this model, we determine contrast (flux) ratios for a sample of exoplanets visible during an Antarctic LDB flight. In Section \ref{sec:syserr_corr}, we examine the most important correlated noise sources we will encounter while observing from a balloon platform. These include pointing jitter, field stop losses, sky and telescope emission and transmission variations, detector effects, spectral calibration, and stellar variability. Finally, in Section \ref{sec:sim}, we discuss the implementation and results of a full end-to-end simulation of a EXCITE instrument that accounts for all the noise sources discussed above. We compare the results of the simulation to our radiometric estimates and determine how long we can integrate before correlated noise sources become important. We show that an instrument like EXCITE is photon noise limited for all reasonable integration times. In the analytic calculations, we perform noise estimates of the brightest and dimmest targets shown in Figure \ref{fig:EXCITE_transmittance}. This allows us to constrain expected performance. For the end-to-end simulation, we perform simulations on a typical bright target, representing EXCITE's expected performance during an Antarctic LDB flight. 

\subsection{Uncorrelated systematic errors}\label{sec:syserr_uncorr}

In the following section we consider uncorrelated systematic error due to photon arrival statistics (Section \ref{sec:photnoise}) and the detector (Section \ref{sec:detectnoise}). We express the results as a contrast ratio, which is defined as the noise from a given source divided by the signal from the host star. This enables direct comparison with the expected eclipse depths shown in Figure \ref{fig:EXCITE_transmittance}.

\subsubsection{Photon noise}\label{sec:photnoise}

We calculate the photon noise contribution from the target star, the Earth's atmosphere, and the telescope optics. For a given source photon flux $\phi_{\gamma}\!\left(\lambda\right)$, we calculate a corresponding electron flux at the detector $\phi_{e^{-}}\!\left(\lambda\right)$: 

\begin{equation}
\phi_{e^{-}}\!\left(\lambda\right) = \eta\!\left(\lambda\right) \times \phi_{\gamma}\!\left(\lambda\right),
\end{equation}
where $\eta\!\left(\lambda\right)$ is a generalized efficiency term and $\lambda$ indicates that each term is wavelength-dependent. The noise variance in the electron flux $\varepsilon_{e^{-}}\!\left(\lambda\right)$ is given by the square root of the total electron flux:
\begin{equation}
\varepsilon_{e^{-}}\!\left(\lambda\right)=\sqrt{\phi_{e^{-}}\!\left(\lambda\right)}.
\end{equation}
From here onward, we drop the $\lambda$ for clarity. 

For the target star, the phonon flux $\phi_{\gamma}^{\mathrm{star}}$ is calculated from its blackbody emission, and $\eta^{\mathrm{star}}$ is given by the product of atmospheric transmittance, telescope optical efficiency, and detector quantum efficiency. Atmospheric transmittance is calculated with MODTRAN and is $>95\%$ over most of the EXCITE band (see Figure \ref{fig:EXCITE_transmittance}). Telescope optical efficiency combines the reflectance of reflecting optics with the transmittance of refracting optics and dichroics. We conservatively take this to be 60\%, assuming that reflecting optics have 3\% emissivity and dichroics, prisms, and the cryostat window can achieve 85\% transmittance. Detector quantum efficiency is based on the specification sheet from Teledyne and is confirmed by measurements. Across the EXCITE band, the detector quantum efficiency exceeds 85\%.

For the Earth's atmosphere, the emission $\phi_{\gamma}^{\mathrm{sky}}$ is calculated using MODTRAN, assuming an altitude of 38 km, an elevation of 45$^{\circ}$ and an azimuth anti-Sun. In reality, the mean altitude in flight will be higher ($>39$ km), so we effectively overestimate the noise contribution from the sky. The efficiency term $\eta^{\mathrm{sky}}$ is given by the product of the telescope optical efficency (which includes a 10\% obscuration from the secondary mirror) and the detector quantum efficiency (see above).

The telescope emission $\phi_{\gamma}^{\mathrm{optics}}$ is dominated by the warm mirrors and warm dichroics. We assume the mirrors and dichroics have an emissivity of 3\% and a temperature of 253 K. Their emission spectra are calculated as grey bodies. The efficiency term $\eta$ is different for each of the warm optical surfaces. Emission from the primary mirror is attenuated (absorbed) by the secondary mirror, the tip/tilt mirror, both warm dichroics, the cryostat window, and all the cold spectrometer surfaces. Emission from the secondary mirror is similar, except it does not self-attenuate; the same is true for the other surfaces. Thus in our model each successive warm source contributes 3.1\% more flux than the previous. The cold surfaces inside the cryostat contribute negligible flux.

\subsubsection{Detector and readout noise}\label{sec:detectnoise}

The detector contributes to measurement noise through its dark current and readout noise. Dark current noise is negligible across the EXCITE bandwidth; noise from other sources dominates by several orders of magnitude. Readout noise is controlled by windowing the array (see Section \ref{sec:FPA}) and by allowing pixels that sample wavelengths longer than $\sim 4$ $\mu$m to saturate. At these wavelengths the sensitivity is already limited by emission from the telescope, field stop losses, and starting at $\sim 4.3$ $\mu$m, emission from the Earth's atmosphere.

\subsubsection{Results}

\begin{figure}
\centering
\includegraphics[width=0.46\textwidth]{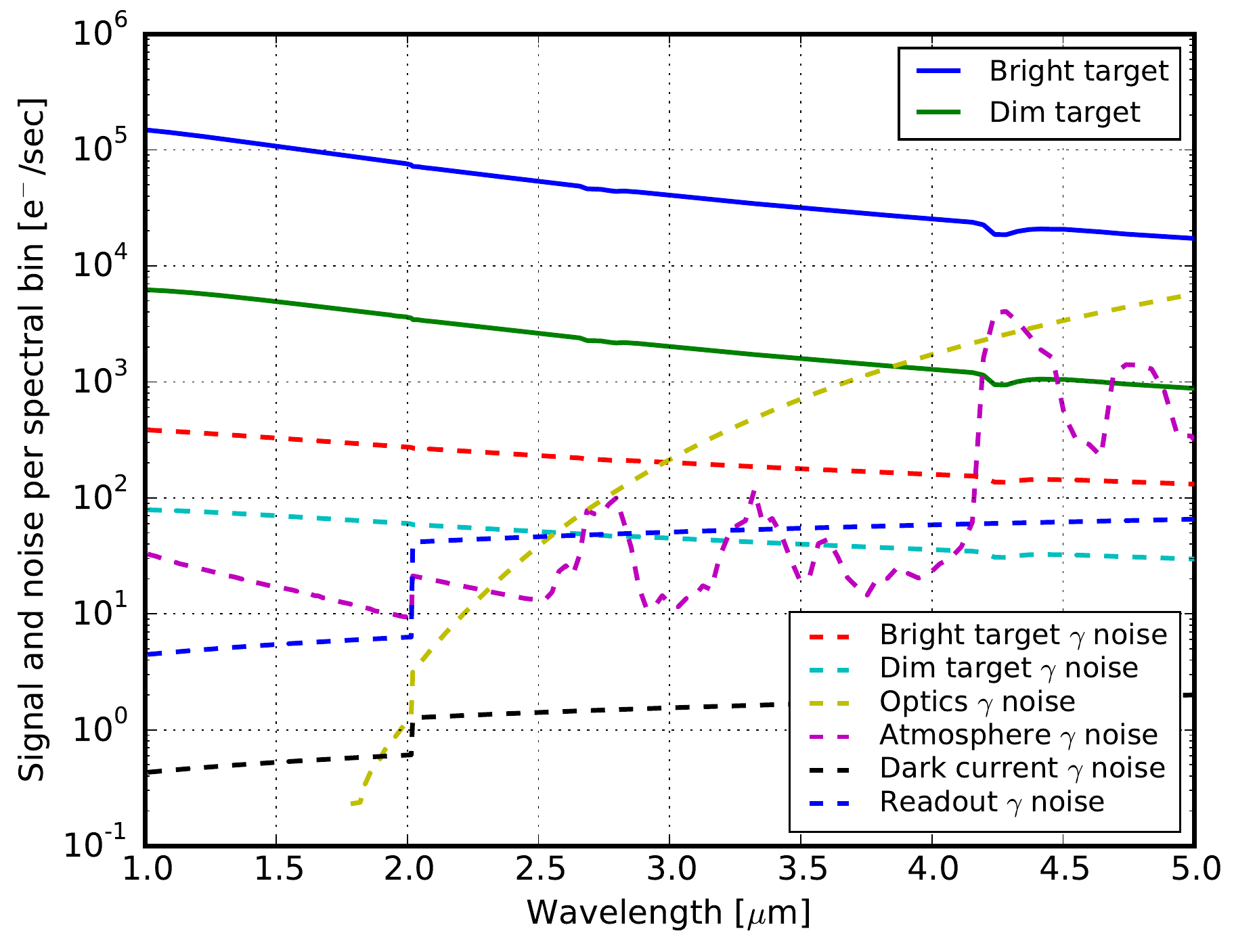}
\includegraphics[width=0.47\textwidth]{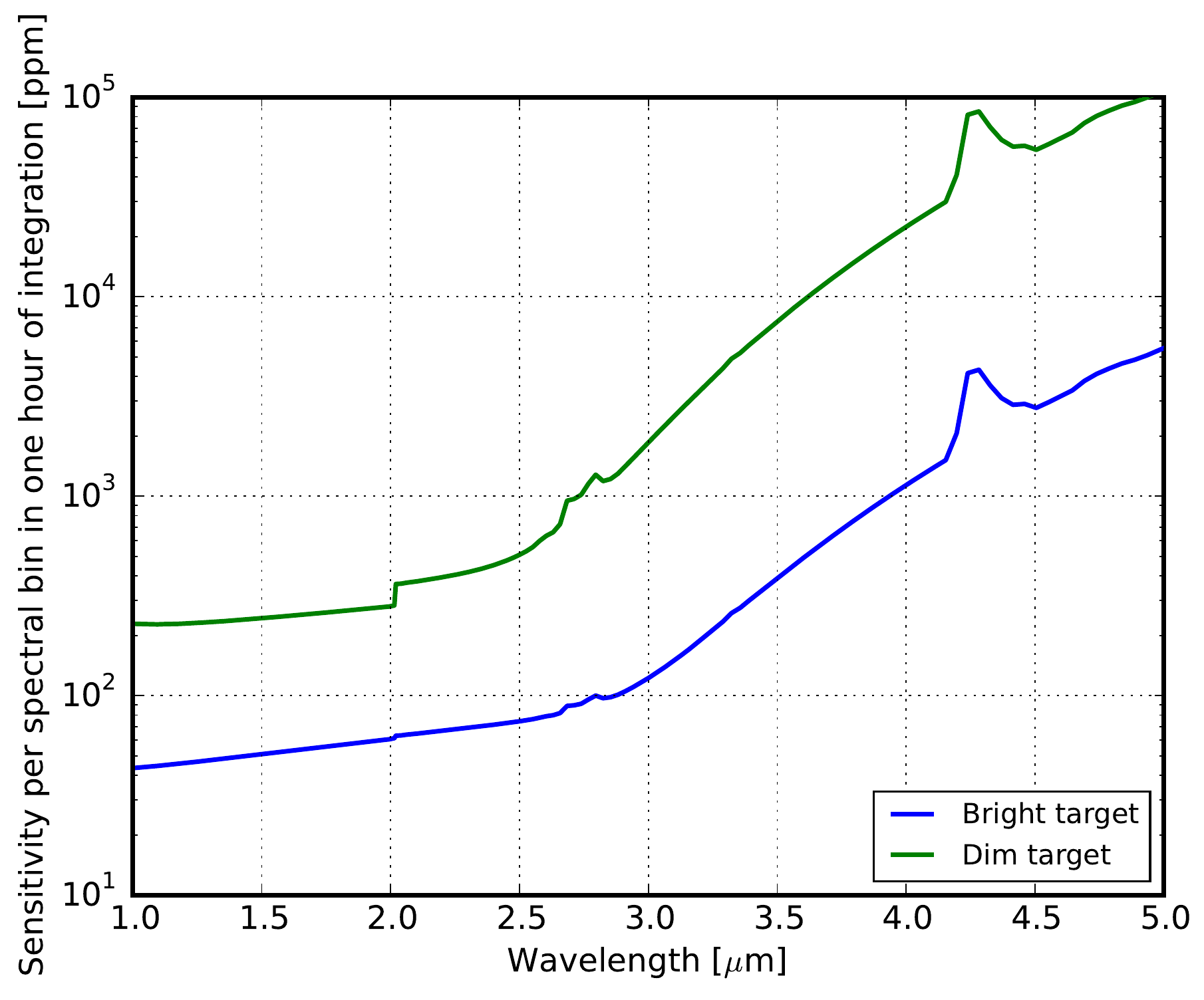}
\caption{Left: Expected signal and 1-$\sigma$ uncorrelated noise levels in one second of integration from the brightest and dimmest targets when binning to $R=50$ (Rayleigh criterion). The discontinuity at a wavelength of 2 $\mu$m indicates the transition between the instrument's spectroscopic channels. Right: Instrument sensitivity per spectral bin at $R = 50$ for the faintest and brightest sources in the target list in one hour of integration. Because the PSF is Nyquist sampled, we are free to re-bin both spectrally and temporally to achieve better sensitivity.}
\label{fig:noise_uncorr}
\end{figure}

After calculating the electron flux from each source, we determine the net photon noise by summing the flux from all sources and taking the square root of the result. We summarize the results in Figure \ref{fig:noise_uncorr}. The left panel shows the expected signal and noise components in one second of integration for the brightest and faintest targets visible from an Antarctic LDB flight at a spectral resolving power of $R = \lambda/\Delta\lambda = 50$ (Rayleigh criterion). The right panel of Figure \ref{fig:noise_uncorr} shows the instrument's sensitivity (noise divided by signal) observing the same targets and integrating for an hour. For observations of targets with intermediate brightness, EXCITE will achieve sensitivity in between these upper and lower bounds. Comparing the results of the radiometric model to the expected K-band eclipse depths of the target sample (Figure \ref{fig:EXCITE_transmittance}), EXCITE has adequate sensitivity to measure eclipse depths with high SNR in one hour of integration. For a typical hot Jupiter orbital period of $\sim 2$ days, this corresponds to phase resolution of $7.5^{\circ}$. As long as uncorrelated noise dominates, sensitivity can be improved by longer temporal and wider spectral binning.

\subsection{Correlated systematic effects\label{sec:syserr_corr}}

The noise budget and sensitivity estimates displayed in Figure \ref{fig:noise_uncorr} assume only uncorrelated photon noise and detector noise. For a realistic instrument, we must determine the timescales at which correlated noise becomes important. In this section, we discuss the contributions from correlated systematic effects we expect to encounter during exoplanet observations from a LDB platform. In particular we consider pointing jitter, field stop losses, atmospheric variations, instrumental drifts, detector variations, observation interruptions, relative spectral calibration, and stellar variability. We assess each effect's impact on EXCITE's photometric stability and the mitigation strategies we will implement at the instrument level and in data analysis. These effects are then included in the end-to-end simulation described in Section \ref{sec:sim}.

\subsubsection{Pointing jitter}\label{sec:jitter}

\begin{figure}
\centering
\includegraphics[width=0.48\textwidth]{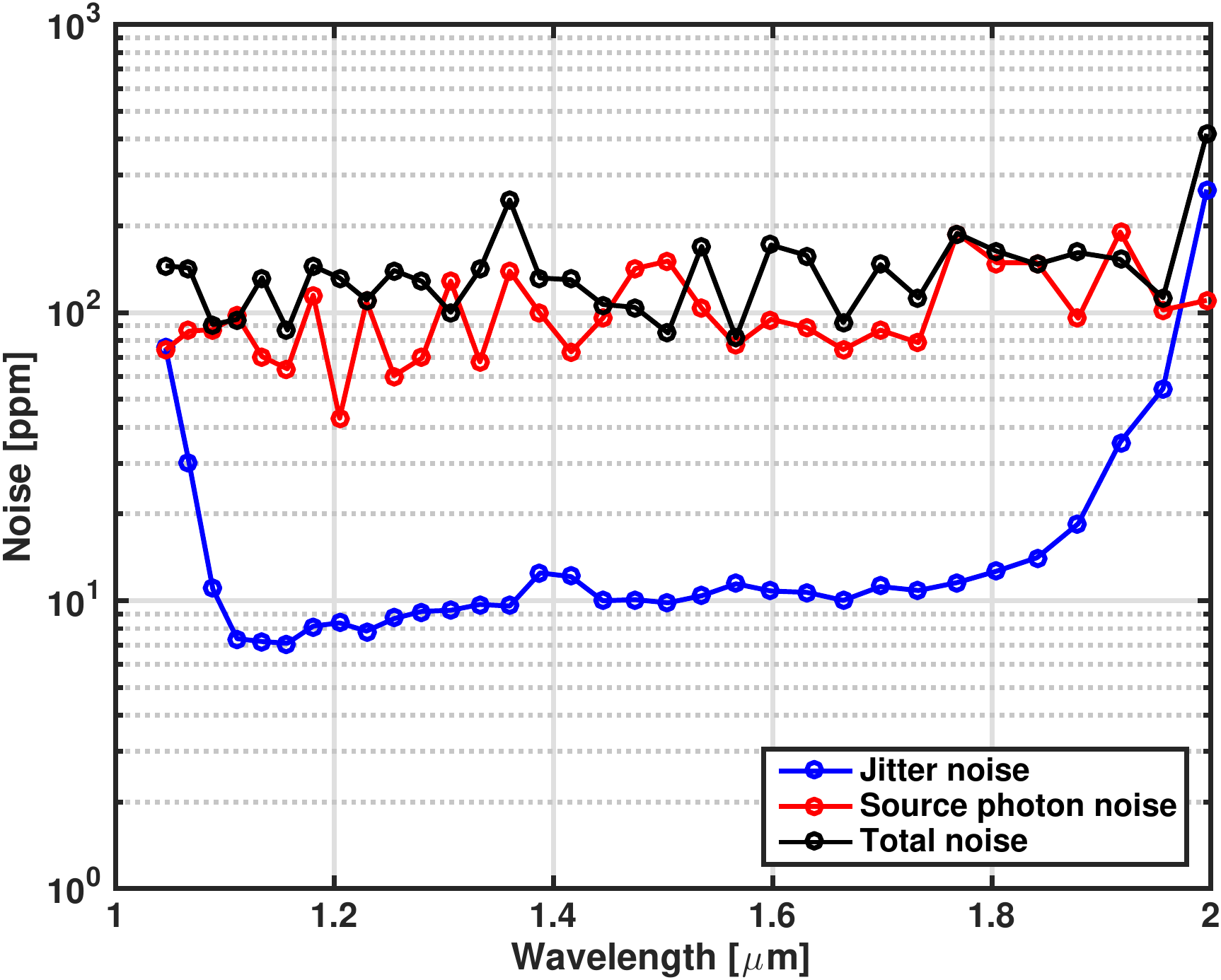}
\includegraphics[width=0.48\textwidth]{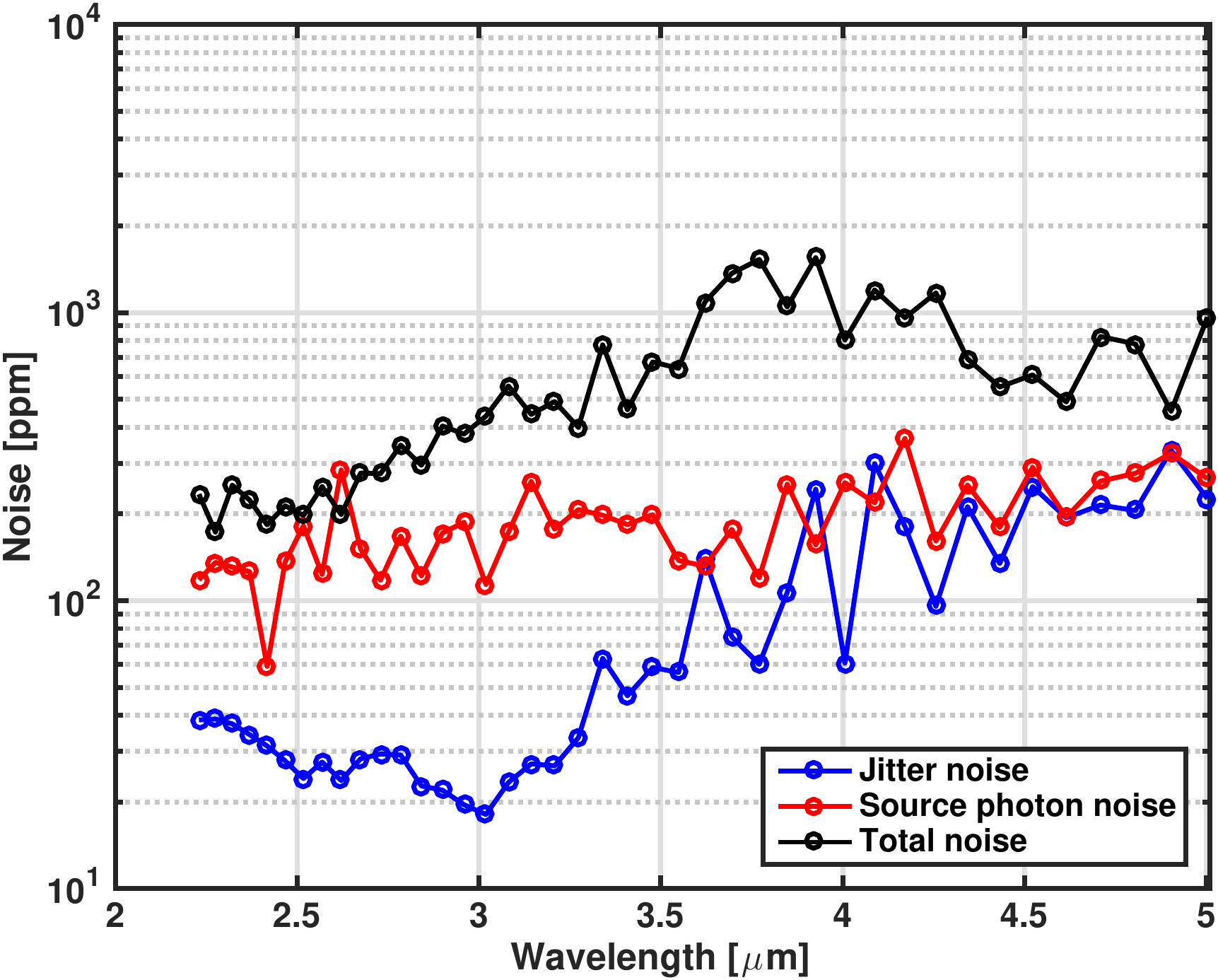}
\caption{Contributions to the noise due to pointing jitter in channel 1 (left) and channel 2 (right) for a bright target in one hour of integration. Jitter was decorrelated using a spectral cross correlation technique and no aperture mask was used. Across the useful science bandpass (1\,--\,4 $\mu$m), jitter noise is roughly an order of magnitude below the source photon noise. The jitter noise becomes comparable to the source photon noise at long wavelengths ($\lambda > 4\,\mu$m) when the contrast between the source and the warm optics and atmosphere becomes negligible (see Figure \ref{fig:noise_uncorr}).} 
\label{fig:jitter_noise}
\end{figure}

Pointing jitter is a source of photometric uncertainty arising from the motion of the sampled spectrum in both the spatial and spectral directions on the focal plane. A spectral image is formed at the detector focal plane array and it is sampled by pixels that have a quantum efficiency that changes across the detector (the inter-pixel response). Each pixel is expected to have a spatial response that deviates from the ideal flat response (the intra-pixel response) and accounts for gaps between adjacent detector pixels.  A jitter in the line-of-sight causes a random motion of the spectral image on the focal plane. The combination of intra- and inter-pixel responses causes a signal modulation that, left uncorrected, can result in a significant source of photometric uncertainty. However, i) intra-pixel responses can be measured (flat field calibration); and ii) if the spectral image is spatially Nyquist sampled, as is the case of EXCITE, no information is lost. 

Our end-to-end time-domain simulations (Section \ref{sec:sim}) study the effect by ``jittering'' the line-of-sight in both the pitch and yaw telescope axes, generating random angular displacements as a function of time with a power spectral density representative of the BIT platform, and including a 5\% intra-pixel response and an inter-pixel response model adequate for the baseline detector array \citep{Barron2007, Pascale2015}. The displacement of the line-of-sight results in a motion of the spectral image in both dispersion and cross-dispersion directions in the focal plane. 

The resulting spectral images are reduced by applying a flat field correction that assumes a 0.5\% knowledge of the flat-field coefficients. To remove the intra-pixel sampling effect we first estimate the shift one spectral image has relative to the first acquired, by cross-correlating the two spectral images. Then all spectral images are shifted onto a common grid by applying a phase in (spatial) frequency space. This shifts the images without affecting their information content.  

In Figure \ref{fig:jitter_noise}, we show the results of simulations of pointing jitter for an effective worst case scenario, where the plate scale is half its actual value (resulting in jitter of 0.6 pixels RMS instead of the expected 0.3 pixels RMS). The results of the simulation show that jitter--induced systematics  are roughly an order of magnitude below the photon noise across EXCITE's science band (1\,--\,4 $\mu$m). This remains true provided that the instrument monochromatic PSF is sampled by at least 2 detector pixels per FWHM ({\sl i.e.}, Nyquist sampled) and the array can be flat-fielded to $\sim0.5$ \%. Nyquist sampling the PSF avoids the most important correlated systematic effects in {\sl Spitzer}~\cite[{\sl e.g.},][]{Ingallsetal2012, Demingetal2014, Pontetal2006}. 

Using the calibration source (Section \ref{sec:FPA}), we can achieve flat field accuracy of $0.5\%$ during ground characterization of the array. Characterizing the array under the same operating conditions as flight, the ground flat field maps will apply to flight data. Additional verification is achieved by observing sky background emission during ascent and at float, and the internal calibrator at float.  The {\sl Spitzer} mission was able to achieve flat field accuracy of $\sim 0.1\%$~\cite{IRAChb} by observing diffuse zodiacal emission. EXCITE can achieve similar performance.  

\subsubsection{Field stop losses}\label{sec:FSL}

\begin{figure}
\centering
\includegraphics[width=1.0\textwidth]{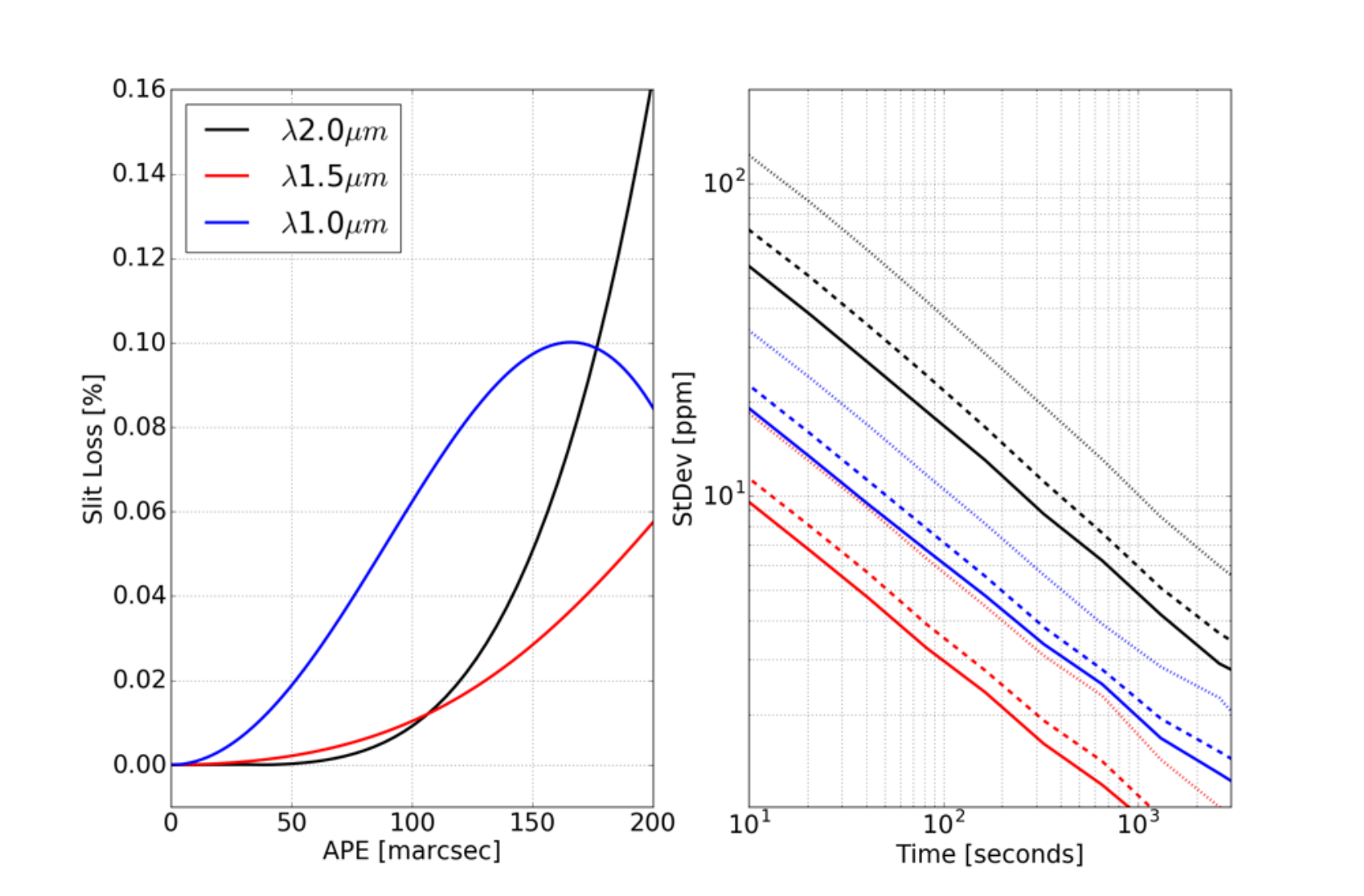}
\caption{Effect of field stop losses on photometric stability in channel 1. Left: field stop loss versus absolute pointing error at three different wavelengths. The jitter of the line of sight modulates the star signal along the curve corresponding to a given wavelength, and around the value corresponding to a given absolute pointing error. Right: photometric error versus integration time for an absolute pointing error of 0\arcsec (solid curves) 0.050\arcsec (dashed curves), and 0.100\arcsec (dotted curves). The results for channel 2 are the same (except the blue, red, and black curves correspond to wavelengths of 2 $\mu$n, 3 $\mu$m, and 4 $\mu$m, respectively.)}
\label{fig:slit_loss}
\end{figure}

Field stop losses are photometric errors arising from both random point errors (RPE) and absolute pointing errors (APE). Field stop losses occur when the PSF is vignetted by the field stop, reducing the efficiency of the instrument. We simulate the effect in the time domain using an RPE of 0.100\arcsec (twice that expected from the fine pointing control) and a BIT jitter power spectrum. The results are shown in Figure \ref{fig:slit_loss}. We find that the integrated effect contributes less than 10 ppm to the noise budget over 1 hour for worst case RPE and APE. The EXCITE field stops are chosen to be the same width as the Airy diameter at the red end of each channel. As a result, APE losses are highest at the blue end of the channel where the secondary maximum of the PSF is vignetted by the field stop. Even though losses at the 10 ppm level are neglible in the EXCITE error budget, performance could be improved by finding an optimal field stop width that reduces modulation of the higher harmonics at short wavelengths. The field stop design employed by EXCITE is similar to that of the {\sl ARIEL} IR spectrometer, where field stop losses were considered and the same conclusion was reached~\cite{arielyb}. 

\subsubsection{Atmospheric variations}\label{sec:atmvar}

The Earth's atmosphere is a non-negligible diffuse background that changes as a function of elevation, altitude and Sun position. Similar variations occur in transmission, though at levels which are at all times subdominant to photon noise. This effect is modeled using MODTRAN and will be validated in-flight by observing calibration stars at different elevations and altitudes. Using MODTRAN, we simulated emission variations based on the altitude profile of the 2006 flight of BLAST \cite{Truchetal2009}. We account for two separate variations. On short timescales, the balloon's altitude oscillates with an amplitude of $\sim 50$ m and a period of $\sim 5$ minutes, resulting in a $\sim 1.5\%$ modulation of atmospheric emission across the EXCITE band. On longer timescales, the balloon altitude oscillates with an amplitude of $\sim 1$ km and a period of $\sim 1$ day, modulating atmospheric emission by $\sim 30\%$. Long-period elevation variations as a target is tracked also modulate sky emission, but at levels an order of magnitude smaller than those from than altitude variations. Integrating for longer than $\sim 5$ minutes removes short-period variations from the simulated time stream. This strategy is consistent with our observing plan, where the shortest observations are $\sim 1$ hour long. Diurnal variations contribute to the noise at a similar level to the uncorrelated sources; however this signal can be monitored and corrected by sampling detector pixels in the cross-dispersion direction and performing background subtraction. 

\subsubsection{Instrumental thermal variations}\label{sec:thermvar}

Instrumental emission will change with time as temperature and gradients change over the structure of the telescope and warm optics. We simulated this effect by allowing the temperature of the warm optics to drift by a few degrees on $\sim 1$ day timescales, matching the performance of BLAST. Noise due to temperature drifts is monitored and corrected in the same way as atmospheric emission. 

Thermal gradients are also likely to introduce throughput variations with time. These are monitored by the wave-front sensing capabilities of the instrument. On the focal plane, thermal gradients modulate the PSF size. PSF stability was measured during the 2016 test flight of BIT \cite{romualdez}, and the results are used to estimate EXCITE's susceptibility to long term drifts. During this flight, BIT achieved $\sim 7\%$ beam solid angle stability over 7.5 minute measurements. During this same interval, the brightest pixel flux was also measured. The beam solid angle stability is nearly anti-correlated with brightest pixel flux measured, but the measurement was not taken with adequate sensitivity to make a determination of the absolute throughput error. Perfect anti-correlation would indicate that the optical efficiency of the instrument is constant. The BIT team expects that the performance of the SuperBIT telescope \cite{SuperBIT2018} will be even better than was achieved by BIT \cite{BIT2019}. EXCITE will use a copy of the SuperBIT telescope. 

\subsubsection{Detector variations}\label{sec:detectvar}

Detector responsivity time variations are small ($< 50$ ppm) on few hour timescales, and can be reduced to below $20$ ppm level by averaging \cite{bezawada2006,clanton2012}. We can reach the 20 ppm level by monitoring pixels in the cross-dispersion direction. Possible longer timescale drifts will be corrected using stable G-type star observations. Detector latencies or persistence are well understood. It has been shown ({\sl e.g.}, the many studies involving {\sl HST/WFC3}) how these can be effectively dealt with using data analysis techniques ({\sl e.g.}, \citet{tsiaras2016} and references therein) and allowing sufficient settling time, which is built into our observing strategy.
The detector is operated in a linear regime, and residual non-linear effects are negligible.
A 5\% divergence from linear response occurs at full well depth~\cite{blanck2011}, and EXCITE will operate at $<75$\% well depth, similar to observations with {\sl HST/WFC3}~\cite{berta2012}, which employs the same detector technology.

\subsubsection{Interrupted observations}\label{sec:interrupt}

Due to the limited range of motion of flexure bearings, the BIT platform requires that the telescope pointing system be reset every $\sim 8$ hours when tracking a target (see Section \ref{sec:pointing}). This process takes about 30 seconds and causes the telescope to temporarily lose pointing control at the focal plane. We expect this operation to have negligible impact on EXCITE's science observations. The demonstrated APE of BIT (up to 0.10'') is well below the plate scale of a detector pixel (0.75'' in channel 1 and 0.50'' in channel 2), so EXCITE will have the absolute pointing accuracy to position the spectrum on the same set of pixels before and after a reset. Thermoelastic instabilities incurred during the reset could case a line-of-sight offset between the star camera and the focal plane. If large enough, this could result in the spectrum landing on a different set of pixels after the reset. This effect can be corrected in real time because we will know with high accuracy where the spectrum is on the focal plane; stellar spectra have easily-recognizable and strong features in the spectral direction and the spatial direction is unambiguous, as shown in Section \ref{sec:jitter}. All other systematic effects occur on timescales that are much longer than the time it takes to reset, therefore the telescope configuration will not change significantly during an interruption. The combination of excellent pointing accuracy and a Nyquist-sampled PSF makes EXCITE relatively immune to systematic effects incurred due to interruptions.

\subsubsection{Relative spectral calibration} 

Relative spectral calibration between the two channels is obtained by ensuring a spectral overlap in the transmission and reflection spectra of the second (cold) dichroic.

\subsubsection{Stellar variability}\label{sec:stellarvar}

Stellar variability of our targets at optical wavelengths due to star spots and stellar rotation has been characterized using the discovery light curves and is generally found to be $<0.1\%$. Where variability is seen, the amplitudes at optical wavelengths are $\sim 1\%$ and the rotation periods of the stars are found to be 10\,--\,20 days. 
Variability in the  1\,--\,5 $\mu$m range is 3\,--\,4 times less than at optical wavelengths for these solar-like stars \cite{EddySunEarthBook}. The infrared variation is strongly correlated with the optical variation so monitoring of the star at optical wavelengths can be used to remove this noise source from our spectroscopic data, as been demonstrated using {\sl Spitzer} data \cite{2012ApJ...754...22K}. We will do the same for EXCITE using the focal plane cameras on the balloon and ground-based optical telescopes. Variability due to granulation and pulsations is negligible over the timescales of the phase curve variation \cite{Sarkaretal2018}. Flares are rarely observed in these targets and will be easily identified in the optical monitoring data and can be removed from the data. 

\subsection{End-to-end simulation}\label{sec:sim}

\begin{figure}[!t]
\centering
\includegraphics[width=0.49\textwidth]{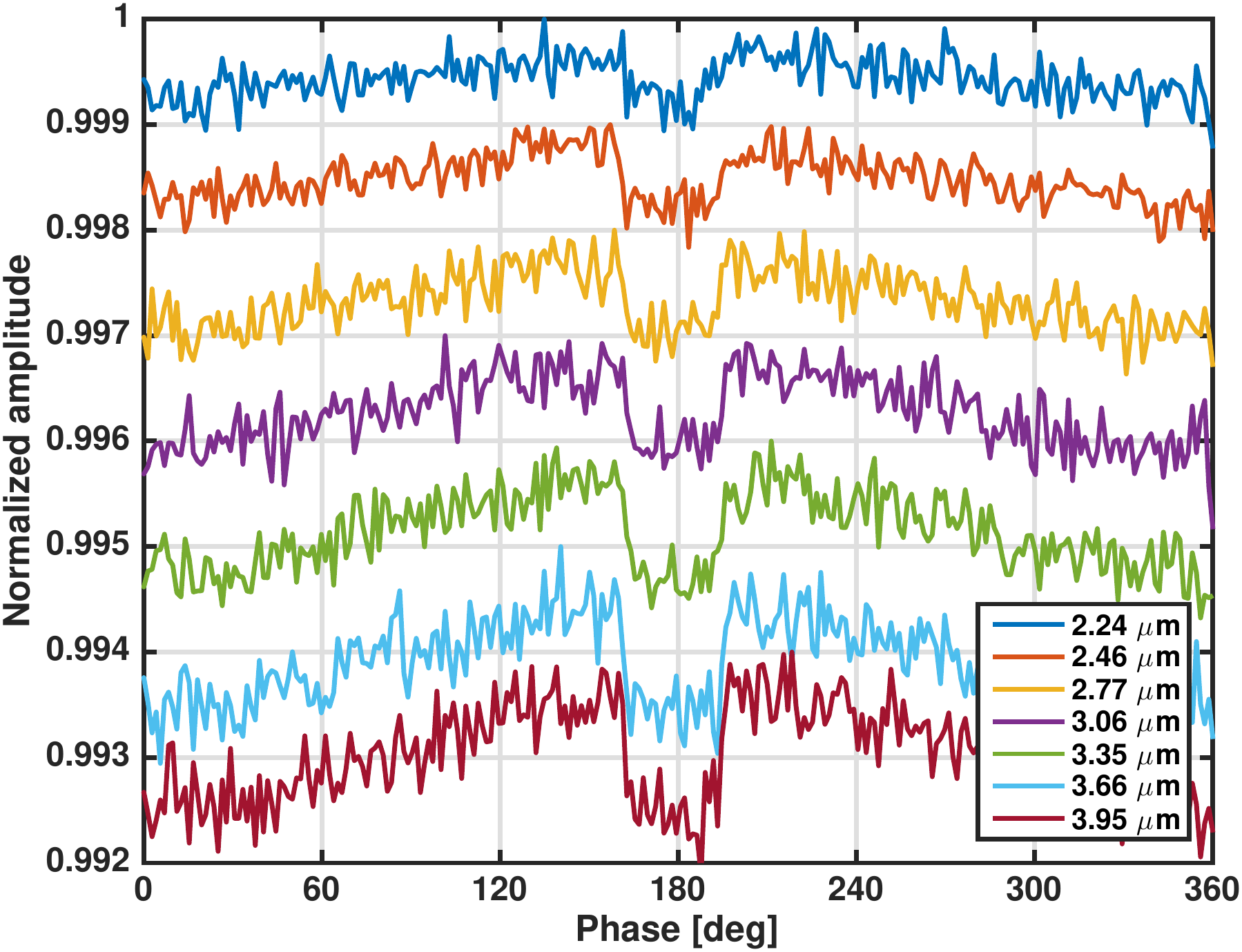}
\includegraphics[width=0.48\textwidth]{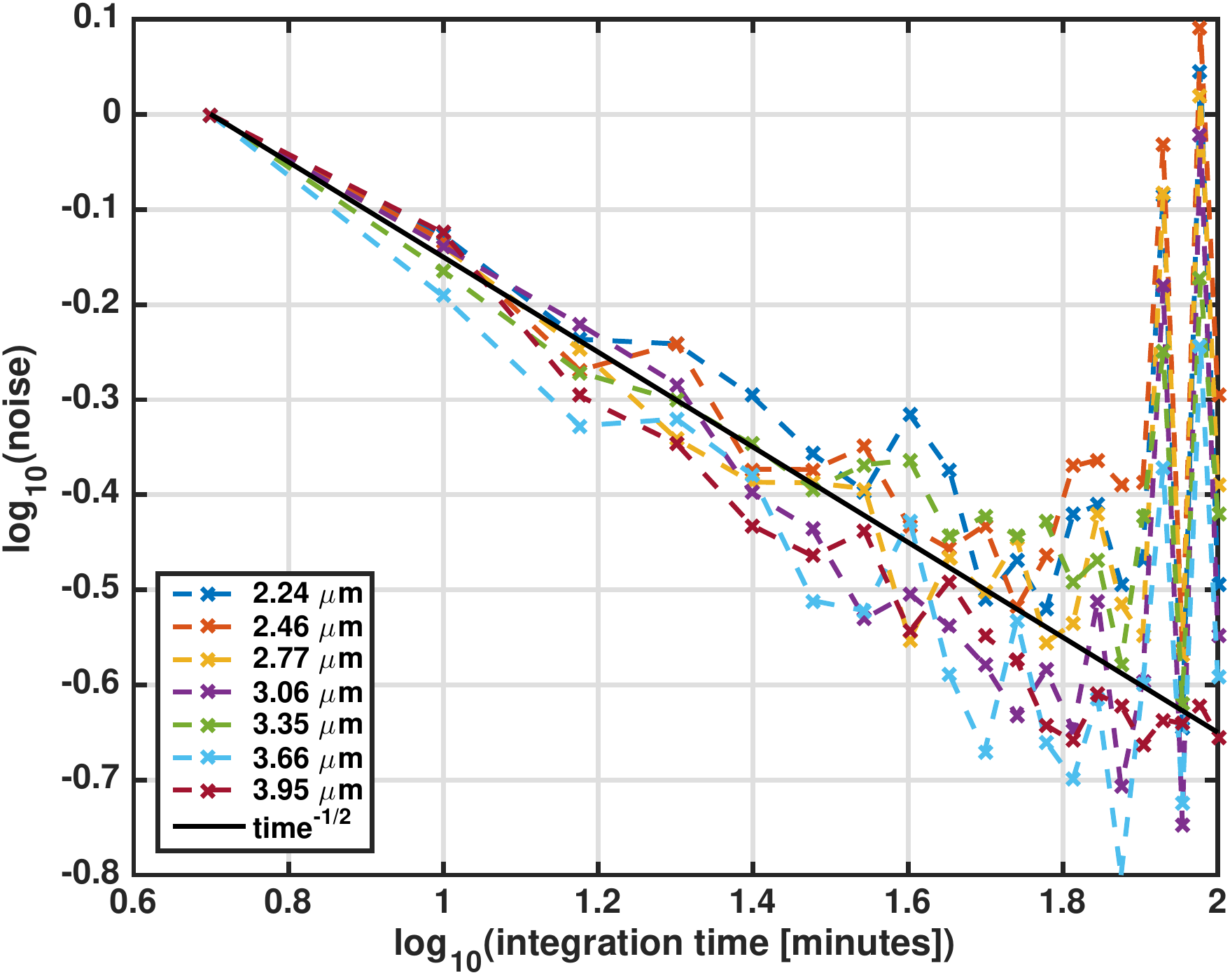}
\caption{Results of the EXCITE end-to-end simulation for an observation of the phase curve of a target of intermediate brightness. Left: recovered phase curves in channel 2 binned to spectral resolution $R \sim 25$ at the long end of the band and using 5 minute exposures. The curves are offset for clarity. The curve corresponding to each wavelength is independently normalized.  The transit is not included in the simulation. The eclipse is centered at a phase of 180 degrees. Right: Measurement noise across channel 2 as a function of integration time for the recovered phase curves (channel 1, with higher SNR, is even more immune to correlated effects). Correlated effects do not contribute to the noise for measurements shorter than $\sim90$ minutes. The large deviations from Poisson noise at long integrations are the result of insufficient temporal sampling of the phase curve.}
\label{fig:err_vs_int}
\end{figure}

To evaluate the effects of correlated systematic effects on EXCITE observations, we developed an end-to-end simulator for EXCITE based on work by \citet{Sarkaretal2016}. The simulation takes into account all the uncorrelated noise sources, as well as correlated noise due to pointing jitter, slit losses, atmospheric variations (from both pointing and altitude drifts), detector variations, and instrument emission variations. The simulator generates a FITS timestream of the image on the focal plane for a given integration time. We chose 5 minute integrations as a balance between temporal resolution and simulation time and file size. The FITS file is then analyzed with a pipeline developed for {\sl HST/WFC3} data by \citet{Kilpatricketal2017}. Each frame was background subtracted by considering a background window of pixels below and above the spectrum spanning the entirety of the dispersion direction.  A median value for each column of the background window was taken to produce a one-dimensional background correction.  The 1-D solution was then smoothed in the dispersion direction to correct for outliers.  The column by column background value was then subtracted from each pixel of the image. The trace and wavelength solutions are provided by the output of the simulation.  The range of wavelengths included in our aperture are divided into 7 bins of width $\sim 0.3$ $\mu$m.  The spectrum from each frame is compared to the spectrum of the first frame using cross correlation in Fourier space to check for any shift in the wavelength--pixel solution.  Each column was summed and weighted by the fraction of that pixel in the bin.  Each eclipse fit was based on the model of \cite{MandelAgol} for a uniform occultation implemented in Python by the BATMAN package \citep{2015BATMAN}.  The orbital, stellar, and planetary parameters sourced from the exoplanets.org database \citep{exodat}  were used as input parameters to the model. 

We first verified that the simulator reproduces the radiometric calculations (Section \ref{sec:syserr_uncorr}). Then we turned on correlated noise sources one by one and compared their simulated magnitudes to analytic estimates. Finally, we included all sources of error. The results of the simulation are shown in Figure \ref{fig:err_vs_int}. The figure shows retrieved spectroscopic phase curves, and how noise in channel 2 depends on integration time for a moderately bright, short orbital period planet. We find that for wavelengths up to 3.95 $\mu$m, and for integrations up to 90 minutes, the simulated noise obeys Poisson statistics. With higher SNR, channel 1 is even more immune to correlated systematic effects. 

\begin{figure}
\centering
\includegraphics[width=1\textwidth]{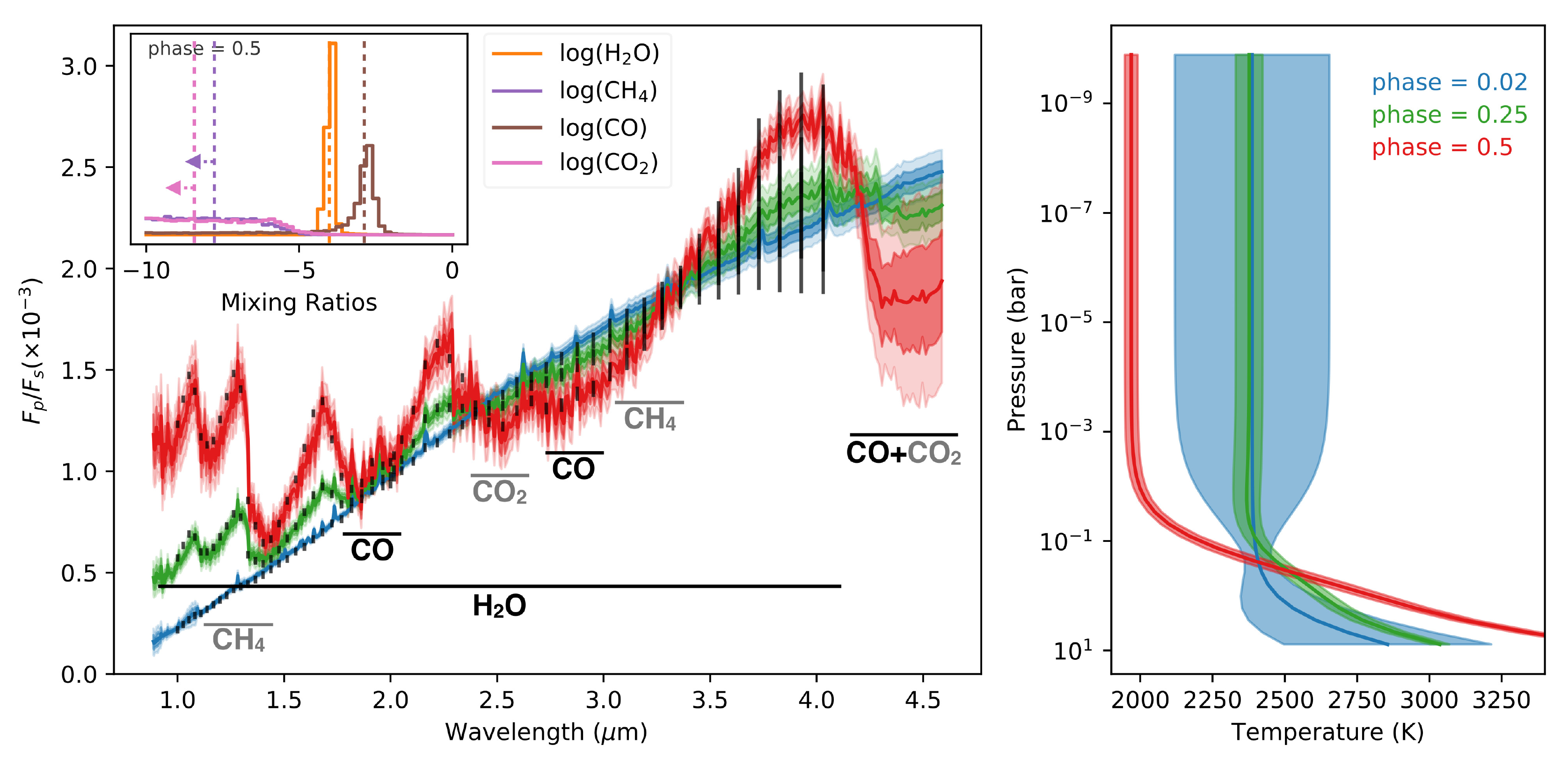}~~~~%
\caption{Spectrocopic retrieval simulations for the phase curve of a moderate brightness, short period hot Jupiter. Using the TauREx retrieval framework, a chemical equilibrium day \& night-side chemistry with C/O = 0.8 and solar metallicity was calculated. We base the day-side temperature profile on \cite{2016arXiv161108608M} and assume an isothermal profile for the night contribution. Left: Resulting emission spectra at phases 0.02 (blue), 0.25 (green) and 0.5 (red). The error bars correspond to calculated noise from the radiometric model, with an additional added noise floor that is equivalent to 50 ppm in an hour of integration. At C/O = 0.8 the main observables are water and carbon monoxide. These could be accurately retrieved with high degree of confidence, dotted lines show input values. At higher C/O ratios CH$_4$, CO$_2$ and other species become increasingly abundant, we labeled their strongest absorption bands in gray. Right: Temperature profiles retrieved for varying phases. As more and more day-side emission becomes visible, the temperature profile departs strongly from the initial isothermal, as expected. The instrumental performance here can be directly compared to the {\sl HST} results shown in Figure \ref{fig:Phasecurve}.}
\label{fig:retrieval}
\end{figure}

With the performance of the instrument well understood, we performed a simulated retrieval of the phase curve of the same sample planet using the TauREx retrieval algorithm \citep{2015ApJ...813...13W,2015ApJ...802..107W}.  The retrieval results are shown in Figure \ref{fig:retrieval}. For this retrieval, we used the radiometrically-calculated instrument noise model binned to $R=20$, with an added noise floor that is equivalent to 50 ppm in an hour. This added noise is more than sufficient to account for the  systematic effects that were studied in Section \ref{sec:syserr_corr}, which we found to contribute 40 ppm in an hour at $R=50$. Additionally, treating the systematic errors as an added noise floor is consistent with the result that the noise obeys Poisson statistics (Figure \ref{fig:err_vs_int}.) Compared to the {\sl HST}-measured spectroscopic phase curve (Figure \ref{fig:Phasecurve}), EXCITE achieves comparable sensitivity at short wavelengths, but critically, has sensitivity over a much broader band (1\,--4\, $\mu$m instead of 1.1\,--\,1.7 $\mu$m.) This enables observations through the peak of the planet's SED, enabling direct constraints of the planet's global energy budget and circulation patterns, and observations where molecular spectra are not lost to dissociation and the presence of H$^{-}$.

\section{Conclusions}\label{sec:conc}

Spectroscopic phase curve observations are rich scientifically but extremely resource intensive. A NIR spectrograph flying on a high-altitude balloon platform is well-suited to make these measurements. In a single LDB flight such an instrument can collect enough quality data to greatly expand our knowledge of hot Jupiter atmospheric physics. 

Key to making spectroscopic phase curve measurements is keeping instrumental systematic effects at levels below the incident photon noise. Using the EXCITE experiment as a worked example, we show that the most important systematic effects we expect to encounter in the NIR from a balloon platform are subdominant to photon noise for integrations up to 90 minutes in length. The effects we studied are summarized in Table \ref{table:syserrall}.

Current and future space-based observatories are unlikely to make enough spectroscopic phase curve observations to significantly expand the physical parameter space of observed planets. We show that an instrument like EXCITE can fulfill this critical role.

\begin{table}[ht]
\centering
\caption{List of dominant systematic effects and mitigation strategy as implemented at  instrument level (does not include data analysis). In determining the residual amplitude we assume $R=50$ and 1 hour integrations. Adding all the systematic effect contributions in quadrature results in a 40 ppm total contribution. For the atmospheric retrieval (Figure \ref{fig:retrieval}), we added a noise floor equivalent to 50 ppm in an hour at $R=20$.}
\begin{tabular}{|l l l l|}
\hline
Systematic effect & Mitigation strategy & Residual amplitude & Paper  \\ 
 &  &  at $\lambda=2.5$ $\mu$m & section \\
\hline\hline

Pointing jitter & Array flat fielding & $<25$ ppm & \ref{sec:jitter} \\

\hline
Field stop losses & Pointing accuracy and stability& $<5$ ppm & \ref{sec:FSL} \\

\hline
Atmospheric variations & Cross-dispersion sampling & $<10$ ppm & \ref{sec:atmvar} \\

\hline
Temperature drifts & Cross-dispersion sampling & $<10$ ppm & \ref{sec:thermvar} \\

\hline
Detector variations & Ground and flight calibration& $<20$ ppm & \ref{sec:detectvar} \\

\hline
Interrupted observations & Pointing accuracy and stability& $<10$ ppm & \ref{sec:interrupt} \\

\hline
Stellar variablility & Optical wavelength monitoring& $<10$ ppm & \ref{sec:stellarvar} \\

\hline
\hline
Combined error & & $<40$ ppm & \\

\hline

\end{tabular}
\label{table:syserrall}
\end{table}

\section*{Acknowledgments}

This project has received funding from the European Research  Council  (ERC)  under  the  European  Union’s Horizon 2020 research and innovation programme (grant agreement No 758892, ExoAI) and under the European Union’s   Seventh   Framework   Programme   (FP7/2007-2013)/  ERC  grant  agreement  numbers  617119  (Exo-Lights). 



\bibliographystyle{ws-jai}
\bibliography{excite}

\end{document}